\documentclass[twoside]{amsart}
\usepackage{url}
\usepackage[latin1]{inputenc}
\usepackage{graphicx}
\usepackage{color}
\usepackage{pstricks}

\usepackage{cite,caption,bbold}

\def\cM{{\mathcal M}}
\def\cL{{\mathcal{L}}}
\def\RR{{\mathbb R}}
\def\thetab{\boldsymbol{\theta}}
\def\omegab{\boldsymbol{\omega}}

\def\alphab{\boldsymbol{\alpha}}
\def\betab{\boldsymbol{\beta}}
\def\gammab{\boldsymbol{\gamma}}

\def\Omegab{\boldsymbol{\Omega}}
\def\dd{\mathbf{d}}

\newcommand{\bo}[1]{\boldsymbol{#1}}

\begin{document}

\bibliographystyle{unsrt}

\title{Discrete differential forms for cosmological space-times}
\author{Ronny Richter}
\email{richter@na.uni-tuebingen.de}
\address{Mathematisches Institut, Universit\"at T\"ubingen,
  Auf der Morgenstelle 10, 72076 T\"ubingen, Germany}
\author{J\"org Frauendiener}
\email{joergf@maths.otago.ac.nz}
\address{Centre of Mathematics for Applications, University of Oslo,
  P.O. Box 1053, Blindern, NO-0316 Oslo, Norway}
\curraddr{Department of Mathematics and Statistics, University of
  Otago, P.O. Box 56, Dunedin 9010, New Zealand}

\begin{abstract}
In this article we describe applications of the numerical method of
discrete differential forms in computational GR. In particular we consider the
initial value problem for vacuum space-times that admit plane gravitational
waves.
As described in an earlier paper the discrete differential form approach can
provide accurate results in spherically symmetric space-times
\cite{Richterfrauendiener2006}. Moreover it is manifestly coordinate
independent.

Here we use the polarised Gowdy solution as a testbed for two numerical
schemes. One scheme reproduces that solution very well, in
particular it is stable for a comparatively long time and converges quadratically.


\end{abstract}
\maketitle

\thispagestyle{empty}
\section{Introduction}
In an earlier paper on the application of discrete differential forms
in numerical General Relativity we described this method for
spherically symmetric problems \cite{Richterfrauendiener2006}. Here we
apply the numerical schemes in the context of $\mathbb T^2$
symmetric cosmological space-times.

In \cite{Richterfrauendiener2006} we used discrete differential forms
to calculate the geometry of the domain of dependence of a given
initial hypersurface. In that context the domain of dependence was
only a small section of the full space-time, which makes it impossible
to investigate the longterm behaviour of the numerical schemes unless
one is prepared to introduce outgoing boundary conditions. However,
this would not allow us to judge the pure evolution algorithm.  Thus,
we consider space-times which have compact Cauchy surfaces,
i.e. cosmological solutions of Einstein's equations.

Thus, we will consider space-times $\cM$ with the topology $\Sigma\times
\mathbb{R}$, where $\Sigma$ is a closed 3-manifold homeomorphic to the
Cauchy surfaces of $\cM$.  The topology of the {Cauchy surface} can be
controlled quite easily within the discrete differential form
approach, because it is given in the very beginning {in terms of}
\emph{incidence matrices} \cite{bossavit1998:_discr_em_prob}.  Hence,
we do not need to worry about outgoing boundary conditions
anymore. For instance, if the topology of $\Sigma$ is the 3-torus
$\mathbb T^3$ then one may equally well use the interpretation that
periodic boundary conditions are imposed in each direction. We will
not consider the full 3-dimensional case here but as
in~\cite{Richterfrauendiener2006} we will confine ourselves to cases
where the problem is effectively $1+1$-dimensional by imposing
appropriate symmetries.

An example for a $\mathbb T^2$ symmetric space-time which admits
Cauchy surfaces with topology $\mathbb T^3$ is the polarised Gowdy
solution \cite{GowdySpaceTime}. This solution can be interpreted as
describing gravitational waves in a cosmological universe. In contrast
to the static spherically symmetric space-times
of~\cite{Richterfrauendiener2006} this solution is time-dependent.
Using this solution to {test the method} has another advantage: it is
one of the testbeds for numerical codes suggested by the Apples with
Apples alliance \cite{Alcubierre:2003pc} and thus one may compare the
results with other numerical methods.

The plan of the article is as follows. In section
\ref{sec:preliminaries} we shortly summarise the properties of
discrete differential forms (more information can be found in
\cite{bossavit1998:_discr_em_prob,frauendiener2006:_discr_differ_forms_gener_relat,Richterfrauendiener2006}).
Then, in section \ref{plane_Sparling_equations} we describe the
equations which result from a symmetry reduction. In section
\ref{discrete_form_section} we present the method of implementing
these equations in two fully discrete numerical schemes. In section
\ref{scenarios} we discuss how the schemes were tested, and in section
\ref{results} we present the results of those tests.

\section{Preliminaries}
\label{sec:preliminaries}

Discrete differential forms in numerical GR were first introduced in
\cite{frauendiener2006:_discr_differ_forms_gener_relat}. The main
feature of this method is its manifest coordinate
independence. Coordinate invariance is one of the fundamental
properties of GR. Thus, it is natural to use coordinate invariant
numerical methods. A discretisation procedure with that property is
Regge calculus \cite{gentle2002:_regge,regge1961:_gener_relat}.
However, so far it did not play a role in numerical GR.  In
computational approaches to treat the problem of coordinate
dependencies multiple coordinate systems are used to cover the
space-time
\cite{Gomez:gr-qc9702002,Thornburg:cqg21_3665,Caltech:gr-qc/0607056,LSU:cqg21_S553},
but there the coordinate invariance is not manifest.

A fundamental difference of the discrete differential form approach in
comparison with many other numerical methods is the discretisation of
space-time. In contrast to the usual procedure, where the finite
counterpart of the manifold is a numerical grid that is composed of a
finite number of points (see e.g. \cite{baumgarte-2003-376} for a
review article about numerical GR), the discrete space-time here also
contains objects of higher dimensionality like curves and
surfaces. The collection of points (nodes), curves (edges), surfaces
(faces) and volumes is a cellular paving
\cite{bossavit1998:_discr_em_prob} and it is called
\emph{computational mesh}. The various elements of a cellular paving
are called \emph{cells} and through incidence relations between the
cells the topology of the computational domain is defined. We use
cellular pavings whose elements are all \emph{simplices}, because this
provides an elegant way to define a discrete version of the exterior
product (see
\cite{frauendiener2006:_discr_differ_forms_gener_relat,Richterfrauendiener2006}).
The cellular paving then corresponds to a \emph{simplicial complex}
\cite{Munkres}.

Based on the cellular paving one discretises a theory that is
formulated in terms of differential forms.  Differential $p$-forms can
be viewed as `the objects which are integrated over $p$-dimensional
submanifolds'. That means they provide maps from $p$-dimensional
submanifolds to the reals. Then \emph{discrete} $p$-forms are maps
that assign a number to every $p$-dimensional cell in the cellular
paving.  They have received some attention since
Bossavit~\cite{bossavit1988:_mixed_whitn} had pointed out that they
correspond to the lowest order mixed finite element spaces defined by
N\'ed\'elec~\cite{nedelec1980:_mixed_r} (see
also~\cite{raviartthomas1977:_mixed_fem}). Finite elements of mixed
type have been used successfully in numerical applications to
electrodynamics,
see~\cite{hiptmair2002:_finit_elem_disc_em,bossavit1998:_comput_elect,bossavit1998:_discr_em_prob}.
In numerical GR the finite element method has been used just recently
e.g. in \cite{PSU:PRD73_044028}.

In order to use this approach one needs to have a formulation of
geometries and, in particular, of GR which uses differential forms.  A
formulation of geometries based on differential forms has been
provided by
\'E.~Cartan~\cite{cartan2001:_rieman_geomet_in_orthog_frame}. The
further step towards a formulation of GR using differential forms has
been carried out by several authors. We mention here the work of
Sparling~\cite{sparling2001:_twist_einst} who has set up an exterior
differential system of equations which is closed if and only if the
vacuum Einstein equations hold.
In~\cite{frauendiener2006:_discr_differ_forms_gener_relat} it is shown
in detail how to set up the discrete formalism based on this exterior
system using the ideas explained above.

In summary, the variables of our proposed discrete formulation will be
the integrals of the differential forms in Sparling's formulation of GR
over submanifolds. In order to get
a finite number of variables we use a finite number of these
submanifolds based on a triangulation of the computational domain.
In this article we describe a simplification of the general formalism which
occurs in space-times with a two-dimensional translational symmetry
(see appendix \ref{app:wave_symm}).


\section{The effective equations}
\label{plane_Sparling_equations}

To obtain the effective equations in the space-times of interest
we start with the Cartan formulation of GR (using exterior forms)
\cite{sparling2001:_twist_einst}. The basic variables in this formalism are
the four 1-forms of a pseudo-orthonormal tetrad $\boldsymbol{\theta}^i$,
$i=0,\ldots,3$ \cite{LandauLifschitz}. With them the metric is
\begin{equation}
g=\boldsymbol{\theta}^0\otimes\boldsymbol{\theta}^0-
\boldsymbol{\theta}^1\otimes\boldsymbol{\theta}^1-
\boldsymbol{\theta}^2\otimes\boldsymbol{\theta}^2-
\boldsymbol{\theta}^3\otimes\boldsymbol{\theta}^3=
\eta_{ik}\boldsymbol{\theta}^i\otimes\boldsymbol{\theta}^k.
\end{equation}
For the description of the connection in this formalism sixteen
1-forms ${\boldsymbol{\omega}^i}_k$, $i,k=0,\ldots,3$ are used. The
connection should be compatible with the metric and torsion free,
which translates to the antisymmetry-requirement and the first
Cartan-equation respectively\footnote{Here and in what follows it is
  understood, that the product of differential forms is the
  anti-symmetrised tensor-product, i.e. the exterior product.}:
\begin{align}
\label{eq:torsion-free}
\eta_{ik}{\boldsymbol{\omega}^k}_j+\eta_{jk}{\boldsymbol{\omega}^k}_i&=0,&
\mathbf{d}\boldsymbol{\theta}^i+{\boldsymbol{\omega}^i}_k
\boldsymbol{\theta}^k&=0.  
\end{align}
Furthermore, the metric should fulfil Einstein's vacuum field equations,
which is equivalent to
\begin{align}
\label{eq:Sparling}
\mathbf{E}_i=0,
\end{align}
where $\mathbf{E}_i$ is the \emph{Einstein~3-form} defined by
(see~\cite{sparling2001:_twist_einst})
\begin{align}
\mathbf{E}_i&=
\frac{1}{2}\varepsilon_{ijkl}{\Omegab^{jk}}\thetab^l,
\end{align}
with the \emph{curvature 2-form}
${\Omegab^j{}_k}=\mathbf
d\bo\omega^j{}_k+\bo\omega^j{}_l\bo\omega^l{}_k$ 
\cite{frankel1997:_physic}.

Since the antisymmetry of the connection 1-forms can easily be imposed,
we are thus interested in the following system of equations
\begin{subequations}
\label{eq:vacuum_equations}
\begin{align}
\label{eq:vacuum_equations_torsion}
\dd\thetab^i+{\omegab^i}_k\thetab^k&=0,\\
\label{eq:vacuum_equations_Einstein}
\mathbf{E}_i&=0.
\end{align}
\end{subequations}
Sparling considered these equations on the frame bundle over the
space-time manifold.  He showed that Einstein's equations are
satisfied if and only if the ideal generated
by~\eqref{eq:vacuum_equations} is a closed differential ideal
\cite{sparling2001:_twist_einst}.

Even though the geometry is fixed by \eqref{eq:vacuum_equations},
there is still the freedom of choosing a gauge, i.e. there are Lorentz
transformations $\Lambda^i{}_k$ of the tetrad that do not change the
metric
\begin{equation}
\label{eq:gauge-freedom}
g=\eta_{ik}\thetab^i \otimes\thetab^k=
\eta_{ik}({\Lambda^i}_j \thetab^j) \otimes ({\Lambda^k}_l \thetab^l)
= (\eta_{ik}\Lambda^i{}_j\Lambda^k{}_l)\,\thetab^j\otimes\thetab^l.
\end{equation}

In this work we will concentrate on general relativistic systems
that occur in the context of space-times with a two-dimensional
translational symmetry.  Essentially that means that the group of
translations $\mathbb R^2$ (or $\mathbb T^2$) acts isometrically
on the space-time and that the orbits of this action are
two-dimensional, space-like, flat submanifolds (the `wave
fronts'). The details are discussed in appendix \ref{app:wave_symm},
where it is also shown how to `factor out' the symmetry action and how
to derive an exterior system on the two-dimensional space
$\cM_1$ of orbits.

This procedure yields two exterior systems. The first one is
\begin{subequations}
\label{eq:pre_min_system}
\begin{align}
\label{eq:pre_min_systema}
\nonumber
\mathbf{d}(f_1+g_1)-(f_0+g_0){\bo\omega}+
(f_0f_1+g_0g_1)\bo\theta^0&\\
+(f_1^2-f_0g_0+f_1g_1+g_1^2)\bo\theta^1&=0,\\
\label{eq:pre_min_systemb}
\nonumber
\mathbf{d}(f_0+g_0)-(f_1+g_1){\bo\omega}+
(f_0f_1+g_0g_1)\bo\theta^1&\\
+(f_0^2-f_1g_1+f_0g_0+g_0^2)\bo\theta^0&=0,\\
\label{eq:pre_min_systemc}
\mathbf{d}{\bo\omega}+
\mathbf{d}(g_0\bo\theta^1+g_1\bo\theta^0)
+(g_0^2-g_1^2)\bo\theta^0\bo\theta^1&=0\\
\label{eq:pre_min_systemd}
\mathbf{d}{\bo\omega}+
\mathbf{d}(f_0\bo\theta^1+f_1\bo\theta^0)
+(f_0^2-f_1^2)\bo\theta^0\bo\theta^1&=0,\\
\label{eq:pre_min_systeme}
\mathbf{d}\bo\theta^0+{\bo\omega}\bo\theta^1&=0,\\
\label{eq:pre_min_systemf}
\mathbf{d}\bo\theta^1+{\bo\omega}\bo\theta^0&=0,\\
\label{eq:pre_min_systemg}
\mathbf d\left(f_0\bo\theta^0+f_1\bo\theta^1\right)&=0.
\end{align}
\end{subequations}
Here $(\thetab^0, \thetab^1)$ is a dyad in the two-dimensional
orbit space $\cM_1$ which carries a Lorentzian metric. The $SO(1,1)$
connection on this space is given by the 1-form
$\omegab^0{}_1=:\omegab$. It is a consequence of the equations above that this
connection is torsion free. The geometric properties of
the orbits are described by the functions $f_0$, $f_1$, $g_0$ and $g_1$.
For details see appendix \ref{app:wave_symm}.

The other exterior system that we shall consider is obtained by a
procedure analogous to the spherically symmetric case
\cite{Richterfrauendiener2006}. For the system
\eqref{eq:pre_min_system} it is performed in
\cite{RichterDissertation}.  One ends up with the following system
\begin{subequations}
\label{eq:grav_wave_first_deg}
\begin{align}
\label{eq:grav_wave_first_degc}
\mathbf{d}(\bo\beta+\bo\delta)+(\bo\alpha+\bo\gamma)(\bo\beta+\bo\delta)&=0,\\
\label{eq:grav_wave_first_degd}
\mathbf{d}\bo\omega+\mathbf{d}\bo\delta
+\bo\gamma\bo\delta&=0,\\
\label{eq:grav_wave_first_dege}
\mathbf{d}\bo\omega+\mathbf{d}\bo\beta
+\bo\alpha\bo\beta&=0,\\
\label{eq:grav_wave_first_degf}
\mathbf{d}\bo\theta^0+\bo\omega\bo\theta^1&=0,\\
\label{eq:grav_wave_first_degg}
\mathbf{d}\bo\theta^1+\bo\omega\bo\theta^0&=0,\\
\label{eq:grav_wave_first_degh}
\mathbf{d}\bo\alpha&=0,\\
\label{eq:grav_wave_first_degi}
\mathbf{d}\bo\gamma&=0,
\end{align}
\end{subequations}
where the 1-forms $\bo\alpha$, $\bo\beta$, $\bo\gamma$ and $\bo\delta$
are defined as 
\begin{align}
\label{eq:def1forms}
\nonumber  \bo\alpha &:= f_0\bo\theta^0+f_1\bo\theta^1,&
  \bo\beta &:= f_1\bo\theta^0+f_0\bo\theta^1,\\
\bo\gamma &:= g_0\bo\theta^0+g_1\bo\theta^1,&
  \bo\delta &:= g_1\bo\theta^0+g_0\bo\theta^1.
\end{align}
To impose these correlation one uses the following algebraic equations
\begin{align}
\label{eq:grav_wave_algebraic}
\star\bo\alpha &= \bo\beta,&
\star\bo\gamma &= \bo\delta,
\end{align}
where $\star$ is the two-dimensional Hodge-operator \cite{frankel1997:_physic}.

The systems \eqref{eq:pre_min_system} and
\eqref{eq:grav_wave_first_deg},\eqref{eq:grav_wave_algebraic}
respectively now serve as the bases to derive numerical schemes. To achieve
that we use the methods described in
\cite{bossavit1998:_discr_em_prob,%
frauendiener2006:_discr_differ_forms_gener_relat,Richterfrauendiener2006}.


\section{Implementation of the discrete equations}
\label{discrete_form_section}
In section \ref{plane_Sparling_equations} and appendix \ref{app:wave_symm} we
derived two systems of equations on the orbit space $\cM_1$. Now we explain
how these systems are discretised and develop numerical schemes.
The first scheme is based on the system
\eqref{eq:pre_min_system}. It can be seen as the analogue of scheme~I in
\cite{Richterfrauendiener2006}. The second scheme is based on
\eqref{eq:grav_wave_first_deg},\eqref{eq:grav_wave_algebraic}
using the same ideas as scheme~III in \cite{Richterfrauendiener2006}.

\subsection{Properties of the simplicial mesh}

The first step in the discretisation procedure is the definition of the
finite counterpart of the manifold. In our case this step is common to
both schemes. As we mentioned in section \ref{sec:preliminaries}, that
structure is a simplicial complex $\mathcal S$, and since we are in two
dimensions that means it is composed of 
simplices of dimensions zero (nodes), one (edges) and two (faces).

Here the simplicial approximation $\mathcal S$ of a subset of the
orbit space $\cM_1$ is generated in analogy to the procedure described
in \cite{Richterfrauendiener2006}. {Thus, the starting point for
  the construction of $\mathcal S$ is an initial hypersurface
  $\mathcal I$ and its discretisation, a 1-dimensional simplicial
  complex $\mathcal C_i$ \cite{Munkres}, which we define so that it
  approximates a space-like curve in $\cM_1$ (see section
  \ref{sec:get_init_vals} for details)}.
The difference here is that the topology of $\mathcal I$
is $S^1$, i.e. that we impose periodic boundary
conditions. The domain of dependence of the initial hypersurface is thus
the full orbit space $\cM_1$.

Having {obtained} $\mathcal C_i$, a very elegant and invariant way
to define the 
positions of the nodes in $\mathcal S\setminus\mathcal C_i$ is to send
lightrays $l_i$ from the nodes in $\mathcal C_i$ and to take the
intersections of these light-like geodesics as those positions.

\begin{figure}
\begin{center}
\input{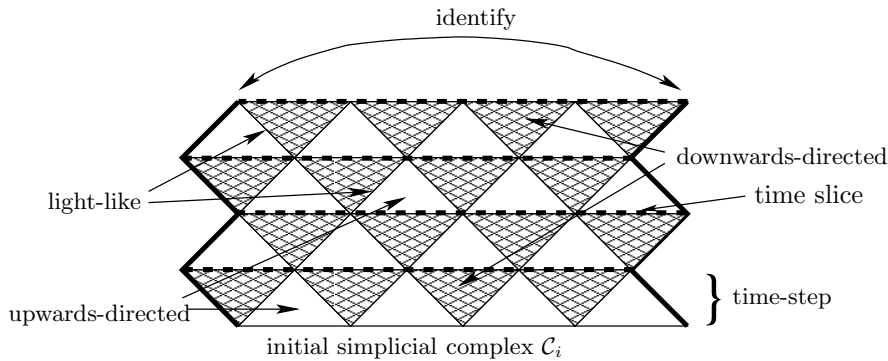}
\caption{The triangulation $\mathcal S$ of the first four time-steps in
the evolution of $\mathcal C_i$.}
\label{fig:triangulation}
\end{center}
\end{figure}

By identifying links between these nodes with the edges of the simplicial
complex appropriately one obtains the simplicial approximation $\mathcal
S$. The construction is described in detail in \cite{Richterfrauendiener2006}
and its result is illustrated in figure \ref{fig:triangulation}.

The complex $\mathcal S$ contains two types of faces, the upwards and the
downwards directed ones. In figure \ref{fig:triangulation} the downwards
directed faces are hatched, and the upwards directed ones are not.
Moreover it contains two types of edges, namely space-like and light-like
ones. Collecting the space-like edges one obtains a structure
that, on the level of topology, appears as $N$ copies of the initial complex
$\mathcal C_i$ (dashed lines in figure \ref{fig:triangulation}).
Propagation from one of these copies (time-slice) to the next one
is one time step.

\begin{figure}
\begin{center}
\input{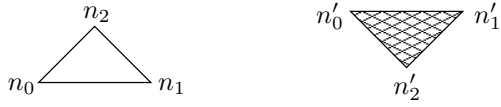}
\caption{The upwards directed faces are denoted $[n_0,n_1,n_2]$ and the
downwards directed faces are $[n_0^\prime,n_1^\prime,n_2^\prime]$.}
\label{fig:faces}
\end{center}
\end{figure}

A $p$-dimensional simplex is commonly denoted $[n_0,\ldots,n_p]$, where
the $n_i$ are the corners of the simplex.
In what follows we will denote the upwards and downwards directed faces
$[n_0,n_1,n_2]$ and $[n_0^\prime,n_1^\prime,n_2^\prime]$ respectively,
such that the edge $[n_0^{(\prime)},n_1^{(\prime)}]$ is space-like and
the edges $[n_1^{(\prime)},n_2^{(\prime)}]$ as well as
$[n_0^{(\prime)},n_2^{(\prime)}]$ are light-like (see figure
\ref{fig:faces}).

As we already discussed in \cite{Richterfrauendiener2006} the construction
through light-like geodesics seems to be the simplest invariant method to
define the position of $n_2$ in $(1+1)$-dimensional manifolds. 
The choice of the nodes at a later time and their connections to the
nodes at the initial time is essentially arbitrary and only restricted
by topological considerations. It is only when the $\thetab^i$ are
known on all the edges that the geometry of the mesh is determined. We
will see later that a part of these values can be specified freely
while the rest is determined from the equations.

As in all numerical simulations degeneracies may occur. For instance two
adjacent nodes in the same time slice may have a time-like distance.
However, this must be seen as a sign that the mesh is too coarse and
should be refined.

\subsection{Properties of discrete forms}
To discretise the equations \eqref{eq:pre_min_system} and
\eqref{eq:grav_wave_first_deg} one needs discrete versions of the operations
exterior product and exterior derivative. If the system
\eqref{eq:grav_wave_algebraic} is taken into account additionally a
discrete Hodge operator must be defined.

To get a discrete exterior product we use the method explained in
\cite{frauendiener2006:_discr_differ_forms_gener_relat} leading to the
following formulas. Let $\bo\alpha^p$ and $\bo\beta^q$ be discrete
$p$- and $q$-forms respectively then
\begin{equation}
\label{eq:ext_prod}
\begin{aligned}
\bo\alpha^0\bo\beta^q[n_0,\ldots,n_q] &=
\frac1{q+1}
\left(\bo\alpha^0[n_0]+\ldots+\bo\alpha^0[n_q]\right)
\bo\beta^q[n_0,\ldots,n_q],\\
\bo\alpha^1\bo\beta^1[n_0,n_1,n_2] &=
\frac1{6}\big(\bo\alpha^1[n_0,n_1]\bo\beta^1[n_0,n_2] -
\bo\alpha^1[n_0,n_2]\bo\beta^1[n_0,n_1] \\
&\qquad+
\bo\alpha^1[n_1,n_2]\bo\beta^1[n_1,n_0] -
\bo\alpha^1[n_1,n_0]\bo\beta^1[n_1,n_2]\\
&\qquad +
\bo\alpha^1[n_2,n_0]\bo\beta^1[n_2,n_1] -
\bo\alpha^1[n_2,n_1]\bo\beta^1[n_2,n_0]\big).
\end{aligned}
\end{equation}

To obtain a natural discrete exterior derivative one applies Stokes' theorem.
For us the relevant formulas are (see
\cite{bossavit1998:_discr_em_prob,frauendiener2006:_discr_differ_forms_gener_relat})
\begin{equation}
\label{eq:ext_deriv}
\begin{aligned}
\mathbf{d}\bo\alpha^0[n_0,n_1] &= \bo\alpha^0[n_1]
-\bo\alpha^0[n_0],\\
\mathbf{d}\bo\alpha^1[n_0,n_1,n_2] &= \bo\alpha^1[n_1,n_2]
-\bo\alpha^1[n_0,n_2] +\bo\alpha^1[n_0,n_1].
\end{aligned}
\end{equation}

The Hodge operator is discretised as described in the context of scheme~III in
\cite{Richterfrauendiener2006}. This discretisation is based on the following
observation.
Let $\bo\alpha$ and $\bo\beta$ be 1-forms such that
$\bo\alpha=\star\bo\beta$. When this equation is {integrated over}
a light-like
geodesic $\gamma_\varepsilon$ with
$\int_{\gamma_\varepsilon}\bo\theta^0=\varepsilon\int_{\gamma_\varepsilon}
\bo\theta^1$ ($\varepsilon\in\{-1,+1\}$) then also  
\begin{align}
\int_{\gamma_\varepsilon}\bo\alpha=\varepsilon\int_{\gamma_\varepsilon}\bo\beta.
\end{align}
Thus we use the following discretisation of the Hodge operator. When the edge
$[n_1,n_2]$ is light-like then
\begin{align}
\label{eq:Hodge}
\star\bo\beta[n_1,n_2]=\varepsilon\bo\alpha[n_1,n_2].
\end{align}
In \eqref{eq:grav_wave_algebraic} the Hodge operator is only applied to
1-forms. Therefore it is not necessary to define its discrete version for
general forms here.

Having a simplicial mesh and the discrete operators we can now develop
the numerical schemes. Common to both schemes is that for each
triangle a system of equations has to be solved.  These are coupled
non-linear algebraic equations. Their analysis is somewhat complicated
and their properties are
not yet clear. They might not have a unique solution.  However, at
least one solution can be found by Newton's iteration method.  We used
the GNU Scientific Library, especially the implementation of the
modified Powell method \cite{powell1970,gsl2005:_refer_manual}.

\subsection{Scheme I}
\label{sec:schemeI}
The first scheme is based on the system \eqref{eq:pre_min_system}. The
variables are the discrete 1-forms $\bo\theta^0$, $\bo\theta^1$ and
$\bo\omega$ as well as the discrete 0-forms $f_0$, $f_1$, $g_0$ and $g_1$.
For the upwards directed faces the numbers
\begin{align}
\nonumber
\{&f_0[n_0],f_0[n_1],f_1[n_0],f_1[n_1],g_0[n_0],g_0[n_1],g_1[n_0],g_1[n_1],\\
&\bo\theta^0[n_0,n_1],\bo\theta^1[n_0,n_1],\bo\omega[n_0,n_1]\}
\end{align}
are given initial data, and
\begin{align}
\nonumber
\{&f_0[n_2],f_1[n_2],g_0[n_2],g_1[n_2],\\
&\bo\theta^0[n_0,n_2],\bo\theta^0[n_1,n_2],
\bo\theta^1[n_0,n_2],\bo\theta^1[n_1,n_2],
\bo\omega[n_0,n_2],\bo\omega[n_1,n_2]\}
\end{align}
are the unknowns.

The system
\eqref{eq:pre_min_system}
is composed of two 1-form equations \eqref{eq:pre_min_systema},
\eqref{eq:pre_min_systemb} and five 2-form equations
\eqref{eq:pre_min_systemc}-\eqref{eq:pre_min_systemg}. {Therefore,
  since there 
  are two 1-form equations on each of the light-like edges and five
  2-form equations on the upwards directed face, it is
clear that the total number of equations in the
discretised system is nine.}
However, we discuss in appendix \ref{app:wave_symm} that
on the continuous manifold the 1-form equations
\eqref{eq:pre_min_systema}, \eqref{eq:pre_min_systemb} take a special
role, because they are not independent
of the remaining equations in \eqref{eq:pre_min_system}.
They are partially redundant.
Therefore we require the discrete 1-form equations
\eqref{eq:pre_min_systema}, \eqref{eq:pre_min_systemb} to be satisfied
at the edge $[n_0,n_2]$, but we do not require these equations at the edge
$[n_1,n_2]$.

Hence, we have seven equations for the upwards directed faces and ten
unknowns. We get three more equations by using the definition of the
position of $n_2$ and fixing the gauge outside the initial
hypersurface. That means we require
\begin{align}
(\bo\theta^0-\bo\theta^1)[n_0,n_2] &= 0,&
(\bo\theta^0+\bo\theta^1)[n_1,n_2] &= 0
\end{align}
to specify the position of $n_2$,
and the gauge is chosen such that
\begin{align}
\bo\omega[n_0,n_2]+\bo\omega[n_1,n_2] = 0.
\end{align}
This gauge choice was also used in scheme III of
\cite{Richterfrauendiener2006}.

For the downwards directed faces the unknowns are
\begin{align}
\{\bo\theta^0[n_0^\prime,n_1^\prime],\bo\theta^1[n_0^\prime,n_1^\prime],
\bo\omega[n_0^\prime,n_1^\prime]\}.
\end{align}
We already discussed how the observation, that the equations
\eqref{eq:pre_min_systema},\eqref{eq:pre_min_systemb}
are redundant, can be implemented on the discrete level.
Following that reasoning we conclude that those 1-form equations need not
be required at $[n_0^\prime,n_1^\prime]$.

It follows that we only need to consider the system
\eqref{eq:pre_min_systemc}-\eqref{eq:pre_min_systemg}.
These are five
2-form equations, but the number of unknowns is only three. Thus, the
system is overdetermined and we have to select three of the five
equations \eqref{eq:pre_min_systemc}-\eqref{eq:pre_min_systemg}.  At
the moment it is not clear how a reasonable choice should be made.
Compared to other choices that were tested shortly and provided
rapidly growing errors the following system seems to work quite well
\begin{subequations}
\label{eq:grav_wave_down_systemI}
\begin{align}
\label{eq:grav_wave_down_systemIa}
0&=\mathbf d\bo\theta^0 + \bo\omega\bo\theta^1,\\
\label{eq:grav_wave_down_systemIb}
0&=\mathbf d\bo\theta^1 + \bo\omega\bo\theta^0,\\
\label{eq:grav_wave_down_systemIc}
0&=\mathbf d\bo\omega + \mathbf d(g_1\bo\theta^0+g_0\bo\theta^1) +
(g_0^2-g_1^2)\bo\theta^0\bo\theta^1.
\end{align}
\end{subequations}

Altogether these are seven equations and seven unknowns for the upwards
directed faces, as well as three equations and three unknowns for the
downwards directed ones.

\subsection{Scheme II}
To obtain the second scheme we discretise the equations
\eqref{eq:grav_wave_first_degc}-\eqref{eq:grav_wave_first_degi},\eqref{eq:grav_wave_algebraic},
i.e.
\begin{equation}
\begin{aligned}
0&=\mathbf{d}\bo\theta^0+{\bo\omega}\bo\theta^1,&
0&=\mathbf{d}\bo\theta^1+{\bo\omega}\bo\theta^0,\\
0&=\mathbf{d}{\bo\omega}+\mathbf{d}\bo\delta
+\bo\gamma\bo\delta,&
0&=\mathbf{d}{\bo\omega}+\mathbf{d}\bo\beta
+\bo\alpha\bo\beta,\\
0&=\mathbf{d}\bo\gamma,&
0&=\mathbf d\bo\alpha ,\\
0&=\mathbf{d}(\bo\beta+\bo\delta)+(\bo\alpha+\bo\gamma)(\bo\beta+\bo\delta),\\
0&=\star\bo\beta - \bo\alpha,&
0&=\star\bo\delta - \bo\gamma.
\end{aligned}\label{eq:red1formsystem}
\end{equation}
That means the variables are the seven 1-forms
$\boldsymbol{\alpha}$, $\boldsymbol{\beta}$, $\bo\gamma$, $\bo\delta$,
$\boldsymbol{\theta}^0$, $\boldsymbol{\theta}^1$ and $\boldsymbol{\omega}$.
The given initial data are hence
\begin{align}
\nonumber 
\{&\bo\alpha[n_0,n_1],\bo\beta[n_0,n_1],\bo\gamma[n_0,n_1],\bo\delta[n_0,n_1],\\
&\bo\theta^0[n_0,n_1],\bo\theta^1[n_0,n_1],\bo\omega[n_0,n_1]\},
\end{align}
and the unknowns for the upwards directed faces are
\begin{align}
\nonumber \{&\bo\alpha[n_0,n_2],\bo\alpha[n_1,n_2],
\bo\beta[n_0,n_2],\bo\beta[n_1,n_2],\\
\nonumber &\bo\gamma[n_0,n_2],\bo\gamma[n_1,n_2],
\bo\delta[n_0,n_2],\bo\delta[n_1,n_2],\\
&\bo\theta^0[n_0,n_2],\bo\theta^0[n_1,n_2],
\bo\theta^1[n_0,n_2],\bo\theta^1[n_1,n_2],
\bo\omega[n_0,n_2],\bo\omega[n_1,n_2]\}.
\end{align}

The system \eqref{eq:red1formsystem} is composed of seven 2-form equations
and two algebraic equations involving the Hodge operator.
These equations are discretised with the exterior product (\ref{eq:ext_prod}), the
exterior derivative (\ref{eq:ext_deriv}) and the Hodge operator
(\ref{eq:Hodge}). For the upwards-directed faces this procedure results in
seven equations for every face and two equations for both of the two
light-like edges. In total these are eleven equations.

With the same procedure as in scheme I, i.e. using that the new edges are
light-like and choosing a gauge with
$\bo\omega[n_0,n_2]+\bo\omega[n_1,n_2]=0$, we reduce the
number of unknowns from fourteen to eleven, such that the number of
equations for the upwards directed faces equals the number of unknowns
there.

For the downwards directed faces the unknowns are the integrals of the
seven 1-forms along the space-like edge, i.e. we have seven unknowns
\begin{align}
\nonumber 
\{&\bo\alpha[n_0^\prime,n_1^\prime],\bo\beta[n_0^\prime,n_1^\prime],
\bo\gamma[n_0^\prime,n_1^\prime],\bo\delta[n_0^\prime,n_1^\prime],\\
&\bo\theta^0[n_0^\prime,n_1^\prime],\bo\theta^1[n_0^\prime,n_1^\prime],
\bo\omega[n_0^\prime,n_1^\prime]\}.
\end{align}

The discrete equations involving the Hodge operator are already satisfied
at the light-like edges $[n_0^\prime,n_2^\prime]$ and
$[n_1^\prime,n_2^\prime]$.
Hence we get seven discrete equations from the 2-forms in
(\ref{eq:red1formsystem}), i.e. from
\eqref{eq:grav_wave_first_degc}-\eqref{eq:grav_wave_first_degi}.


\section{Test-scenarios}
\label{scenarios}
In the last chapters we described, how discrete differential forms can
be applied to study systems with a planar symmetry in GR. In this
chapter we want to present the concrete example, where the code was
tested. The idea for the test is to derive discrete initial data from
an analytical solution, use the numerical schemes to simulate the time
evolution and finally compare the numerical results with the
analytically expected ones.  That means an analytical solution is
needed.

As a testbed we have chosen the polarised Gowdy space-time
\cite{Alcubierre:2003pc,GowdySpaceTime}, because it has at least three
advantages. Its spatial slices have the topology $\mathbb T^3$, the
light-like geodesics take a very simple form and it is suggested as a
testbed for numerical schemes by the Apples with Apples alliance
\cite{Alcubierre:2003pc}.  In standard coordinates $\{t,z,x,y\}$ its
metric reads
\begin{align}
\label{eq:Pol_Gowdy_standard}
g= t^{-1/2}e^{\lambda/2}\left(
\mathbf dt\otimes\mathbf dt-\mathbf dz\otimes\mathbf dz
\right)
-te^P\mathbf dx\otimes\mathbf dx
-te^{-P}\mathbf dy\otimes\mathbf dy,
\end{align}
where $t\in\mathbb{R}$, $z \in [0,1]$,
\begin{align}
\nonumber P(t,z)&=J_0(2\pi t)\cos 2\pi z,\\
\nonumber \lambda(t,z)&=-2\pi t J_0(2\pi t)J_1(2\pi t)\cos^2 2\pi z + 2\pi^2 t^2\left[J_0^2(2\pi t)+J_1^2(2\pi t)\right]\\
\nonumber &\qquad-2\pi^2\left[J_0^2(2\pi)+J_1^2(2\pi)\right]+ \pi J_0(2\pi)J_1(2\pi),
\end{align}
and $J_0$, $J_1$ are the usual Bessel-functions.

It can easily be checked that a light-like geodesic in the $(t,z)$-surface
through a point with 
coordinates $(t_0,z_0)$ also contains
the points $(t_0+k,z_0)$ ($k\in\mathbb N$). In this sense, a massles test particle
needs a time of $\delta t=1$ to travel around the universe. This time interval
is commonly denoted a \emph{crossing time}.

\subsection{The continuous forms}
To get the differential forms we make a gauge-choice, i.e. we
choose some $\bo\theta^0$ and $\bo\theta^1$, that generate the
corresponding metric.
In polarised Gowdy geometry a natural choice leads to
\begin{align}
\label{eq:Polarized_Gowdy_forms}
\nonumber f_0&=\frac{1}{2}t^{\frac{1}{4}}e^{-\frac{\lambda}{4}}\left(\frac{1}{t} +
\partial_tP \right),&
\nonumber f_1&=\frac{1}{2}t^{\frac{1}{4}}e^{-\frac{\lambda}{4}}\partial_zP,\\
\nonumber g_0&=\frac{1}{2}t^{\frac{1}{4}}e^{-\frac{\lambda}{4}}\left(\frac{1}{t} -
\partial_tP\right),&
\nonumber g_1&=-\frac{1}{2}t^{\frac{1}{4}}e^{-\frac{\lambda}{4}}\partial_zP,\\
\nonumber\boldsymbol{\theta}^0 &=t^{-1/4}e^{\lambda/4}dt,&
\nonumber\boldsymbol{\theta}^1 &=t^{-1/4}e^{\lambda/4}dz,\\
\nonumber\boldsymbol{\alpha} &=\frac{1}{2}\left(\partial_z P dz + \partial_t P dt + \frac{1}{t}dt\right),&
\nonumber\boldsymbol{\beta} &=\frac{1}{2}\left(\partial_z P dt + \partial_t P dz + \frac{1}{t}dz\right),\\
\nonumber\boldsymbol{\gamma} &=\frac{1}{2}\left(\frac{1}{t}dt - \partial_t P dt - \partial_z P dz \right),&
\nonumber\boldsymbol{\delta} &=\frac{1}{2}\left(\frac{1}{t}dz - \partial_t P dz - \partial_z P dt \right),\\
{\boldsymbol\omega} &= \frac{1}{4}\left(\partial_z\lambda dt + \partial_t\lambda dz - \frac{1}{t}dz\right),
\end{align}
where for the derivatives of $P$ and $\lambda$ we get
\begin{align}
\nonumber \partial_t P(t,z)&=-2\pi J_1(2\pi t)\cos 2\pi z,\\
\nonumber \partial_z P(t,z)&=-2\pi J_0(2\pi t)\sin 2\pi z,\\
\nonumber \partial_t \lambda(t,z) &= t\left((\partial_t P(t,z))^2+(\partial_z P(t,z))^2\right),\\
\partial_z \lambda(t,z) &= 2t\left(\partial_t P(t,z)\right)\left(\partial_z P(t,z)\right).
\end{align}

\subsection{Association with discrete forms}
\label{sec:get_init_vals}
The next step is to choose an initial hypersurface. This has to
obey our requirement that its topology is $S^1$. That means we
need a closed space-like curve in the two-dimensional orbit space $\cM_1$.%
\footnote{The orbit space $\cM_1$ can be identified with a
surface that satisfies $x=\mbox{const}$, $y=\mbox{const}$.}
For the test we used curves, whose $t$-coordinate is
constant
\begin{equation}
  \label{eq:def_straight_Gowdy}
  \left(
    \begin{array}{c}
      t \\ z
    \end{array}
    \right) =
  \left(
    \begin{array}{c}
      t_0 \\ z_0
    \end{array}
    \right)
    + \lambda   \left(
    \begin{array}{c}
      0 \\ 1
    \end{array}
    \right),
\end{equation}
with $\lambda\in[0,1)$ and fixed $t_0,z_0$.\par

We subdivide this curve into $n$ pieces, which are again of the form
\eqref{eq:def_straight_Gowdy}, but with
$\lambda\in[(i-1)/n,i/n], i=1,\ldots,n$. On each of these pieces
we integrate the 1-forms, and on the nodes we evaluate the 0-forms to get
initial values.

Since light-rays in the polarized Gowdy solution \eqref{eq:Pol_Gowdy_standard}
are the curves with $|dt|=|dz|$, the coordinate of the nodes in the
computational mesh are easily obtained and it is thus straight forward to
compare the numerical and the analytical solution (see
\cite{Richterfrauendiener2006}).

This is done as follows. In every time slice there are $n>1$ space-like edges.
We take the maximum of the relative error of the invariant lengths%
\footnote{We define the relative error to be $|l_{ap}/l_{ex}-1|$, where
$l_{ap}$ is the numerically calculated approximation to the invariant
length and $l_{ex}$ is the length in the exact solution.}
of the edges in a time slice as a measure for the accuracy of the scheme.

In this construction, since $n$ is the number of initial edges,
we need $2n$ time steps to simulate a crossing time ($\delta t = 1$).


\section{Concrete Examples and Results}
\label{results}

Now two scenarios in the polarised Gowdy solution are investigated, an
expanding and a collapsing time evolution. For the expanding evolution
we take an initial hypersurface of the form \eqref{eq:def_straight_Gowdy}
with $t_0=t_0^e:=1$ and $z_0=z_0^e:=0$, and for the collapsing evolution these
constants are $t_0=t_0^c:=9.8753205829098$ and $z_0=z^c_0:=-0.5$.

The hypersurface $t_0=t_0^c$ is in the sense special that the Bessel
function term $J_0(2\pi t)$ vanishes and the Apples with Apples alliance
suggests to use this particular choice.

\subsection{Expanding Evolution}
\begin{figure}
\begin{tabular}{cc}
\includegraphics[scale=0.75]{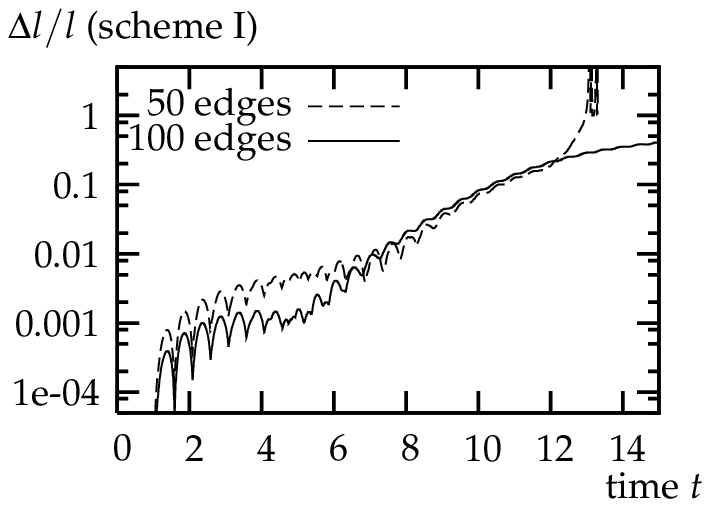}&
\includegraphics[scale=0.75]{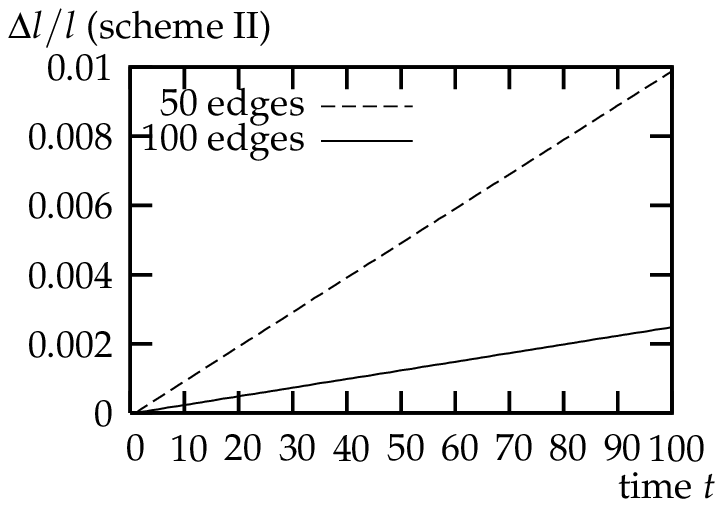}
\end{tabular} 
\caption{Expanding evolution: Growth of the maximal relative error of the
invariant lengths of the space-like edges. Left: Results of scheme I
(logarithmic ordinate). Right: Results of scheme II (linear ordinate).}
\label{fig:gowdy_expand_timeevol}
\end{figure}

In the expanding evolution
we divide the initial curve into $n$ edges of the form
$\{z=\lambda,t=1,\lambda \in [(k-1)/n,k/n],k=1,\ldots,n\}$.
For $n=50$ and $n=100$ the growth in time of the maximal relative error
of the lengths of space-like edges is shown in figure
\ref{fig:gowdy_expand_timeevol}.

For scheme~II
we see that, although the absolute value of the lengths of the edges
grows exponentially in this space-time, the relative error of these lengths
grows only linearly and remains small even at $t=100$. Furthermore in this
scheme the error in the calculation with 100 initial edges is about 4 times
smaller than the error with 50 initial edges. This indicates quadratic
convergence for decreasing edge lengths.

For scheme~I the error does not grow only linearly. We see that for
three to four crossing times the scheme has a good numerical behaviour.
The size of the error grows slowly, it is relatively small and becomes even
smaller when more initial edges are used.
But after four crossing times the error starts to grow exponentially, and
the results of the calculations in different meshes
become similar. Then, at $t\approx 12$ the error in the simulation
with 50 edges blows up rapidly. Similar behaviour can be observed for the
simulation with 100 initial edges, but here the critical time is
$t\approx 20$.

In both schemes the growing of the error is superposed by a quasi-periodical
fluctuation (for scheme~II it is not visible in figure
\ref{fig:gowdy_expand_timeevol}, because it is small, but it exists).
The period of this fluctuation is $\delta t=1/2$, i.e. half as long as the
crossing time.

\subsection{Collapsing Evolution}
\begin{figure}
\begin{tabular}{cc}
\includegraphics[scale=0.75]{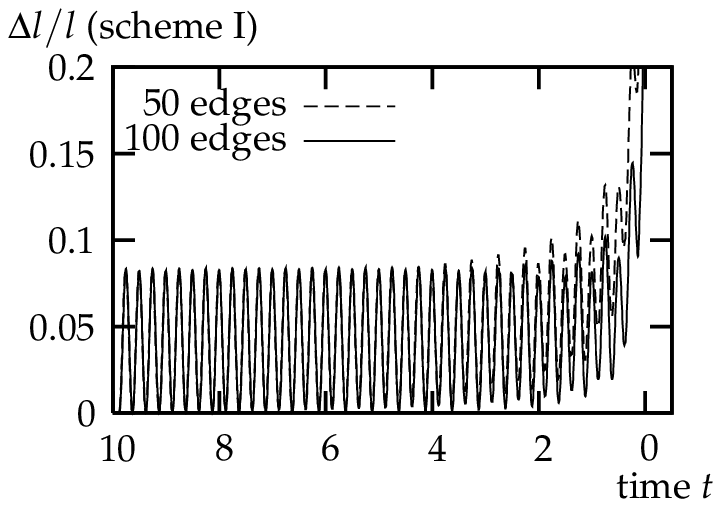}&
\includegraphics[scale=0.75]{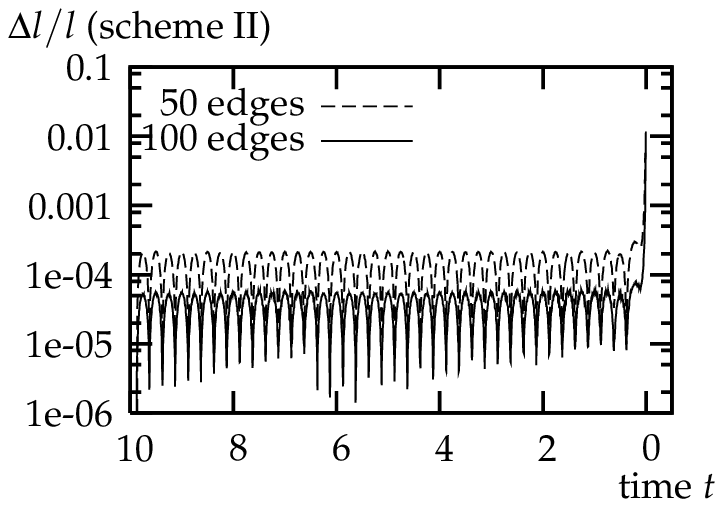}
\end{tabular} 
\caption{Collapsing evolution: Maximal relative error of the invariant lengths
of the space-like edges. Left: Results of scheme~I.
Right: Results of scheme~II.}
\label{fig:gowdy_collaps_time_evol}
\end{figure}

For the collapsing evolution we again
divide the initial curve into $n$ edges of the form
$\{z=\lambda-0.5,t=t_0^c,\lambda \in [(k-1)/n,k/n],k=1,\ldots,n\}$.
For $n=50$ and $n=100$ the growth of the maximal relative error
of the lengths of space-like edges is shown in figure
\ref{fig:gowdy_collaps_time_evol}.

Since at $t=0$ the polarised Gowdy space-time has a singularity (the cosmological singularity), we can only evolve as long as $t$ remains positive. I.e. if we use for instance 50 initial edges, we can at most
make 987 time-steps.

A surprising result is that for both schemes the relative error does not grow at
all. Instead it fluctuates quasi-periodically, but remains at similar sizes
until the time approaches zero, i.e. near the singularity.

Here the period of the fluctuation is $\delta t=1/4$, i.e. a quarter of a
crossing time and only half as long as
in the expanding time evolution. Further investigations reveal that
for other initial hypersurfaces, i.e. $t_0\neq t_0^c$, the error still
fluctuates, but then the period of the fluctuation is $\delta t=1/2$,
like in the expanding time evolution.

Another point to mention is that for scheme~II the error in the simulation
with 100 edges is about four times smaller than the error in the simulation
with 50 edges. This is not the case in scheme~I where the errors of both
simulations are of comparable size.

\subsection{Convergence of the Schemes}

Finally we consider the convergence behaviour of the numerical schemes for
both examples. We saw that in the expanding time evolution scheme~I is not
stable, and that it behaves best for $t<4$. Therefore we investigate the
errors at $t=2$. For the collapsing evolution we choose with $t=t_0^c-4.9$
a time where the (fluctuating) error has neither a minimum nor a maximum.

\begin{figure}
\begin{tabular}{cc}
\includegraphics[scale=0.75]{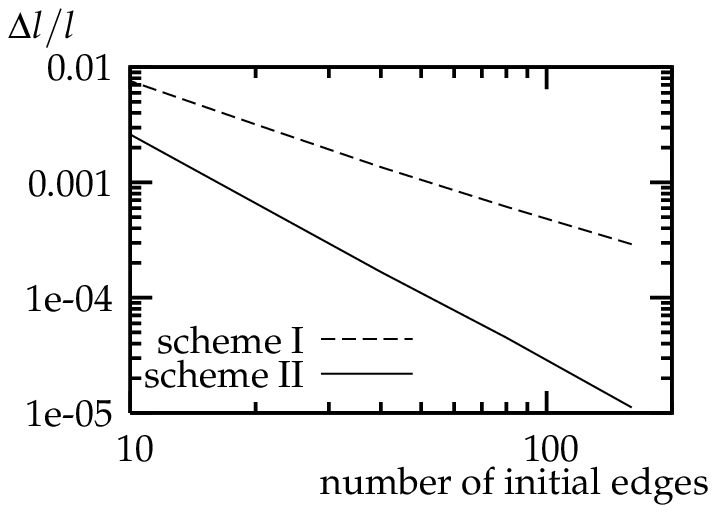}&
\includegraphics[scale=0.75]{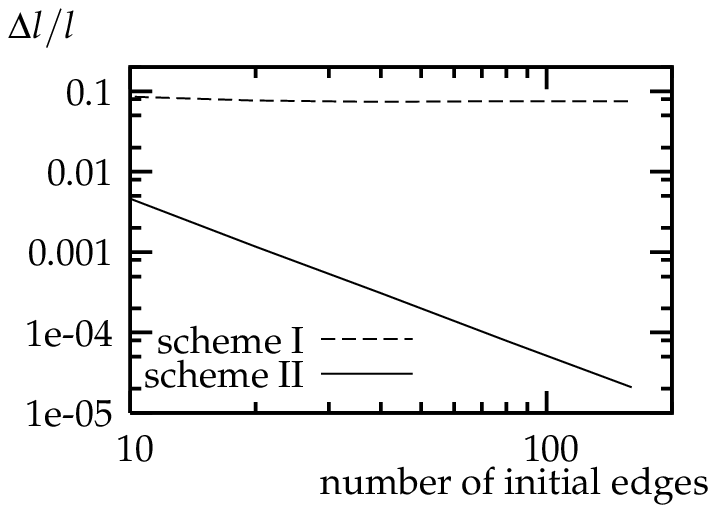}
\end{tabular} 
\caption{Convergence of the schemes. Left: expanding evolution at $t=2$.
Right: collapsing evolution at $t=t_0^c-4.9$.}
\label{fig:gowdy_convergence}
\end{figure}

In figure \ref{fig:gowdy_convergence} it is shown how both schemes behave
when the typical length of the edges is decreased, i.e. when the number
of initial edges becomes larger. To obtain the curves we calculated the
time evolution with 10, 20, 40, 80 and 160 initial edges.

We see that the results of scheme~II converge quadratically to the
analytical solution. The errors of scheme~I on the other hand converge only
linearly to zero in the expanding time evolution and do not converge
at all in the collapsing time evolution. This is in agreement with the
observations that we made in the time evolution simulations.

\subsection{Discussion}

The first point to discuss is the quasi-periodical fluctuation of the
error and its period of $\delta t=1/2$.  This phenomenon points to the
interpretation that, due to the periodic boundary conditions, the
errors that sum up during the time evolution cancel out when a
lightray intersects the other lightray that was sent from the same
point in the other direction.

In the collapsing time evolution we observed that the period of the
fluctuation is only $\delta t=1/4$ when $t_0=t_0^c$, but it is again
$\delta t=1/2$ when other initial data are chosen.  This initial
hypersurface is in the sense special that the Bessel function
$J_0(2\pi t)$ has a root there.  Hence it is most likely that with
these initial data other symmetries of the considered space-time cause
new cancellations of errors.

Concerning the quality of the numerical solutions we see that
(at least for the considered examples) scheme~II provides the best results.
This scheme is accurate and stable, the relative error grows only linearly
in time and remains small. Unfortunately we cannot
say much about its robustness {(in the sense of the Apples with
  Apples alliance)}, because we only considered the polarised
Gowdy space-time. This scheme is the analogue of scheme~III in
\cite{Richterfrauendiener2006}, which also provided the best results
there. {The reason for this behaviour seems to be that the
  system~(\ref{eq:grav_wave_first_deg}) is geometrically preferred.}

The analogue of scheme~I here is scheme~I in
\cite{Richterfrauendiener2006}. For spherically symmetric space-times the
results of that scheme were, compared to the results of scheme~I here,
better. Especially it was quadratically convergent.

For scheme~I here there are situations when it does not converge at all,
and moreover it is not stable. We saw that for small times in the
expanding evolution it behaves quite well, but after that
there is a period when the error grows exponentially. Up to now
it is not clear if the reason for this behaviour is rather the numerical
scheme or the exponentially growing example. However,
it becomes clear that the scheme is unstable when the period
of exponential growth ends (at $t\approx 12$ resp. $t\approx 20$).
We thus conclude that scheme I is not practicable.

Of course it is not too surprising that problems occur in scheme I,
because the system \eqref{eq:pre_min_system}
is not very well understood at the analytical level. In particular,
we were not able to derive a minimal system for the gravitational
wave space-times.
Quite the contrary we found an overdetermined system, and although there
are arguments that some equations are redundant, it was still necessary
to choose discrete equations that we omitted without having justifying arguments
for this particular choice.

Comparing the previously presented numerical schemes with other codes
one has to distinguish between the expanding and the collapsing case.
The reason is that in the collapsing time evolution the coordinate
representation \eqref{eq:Pol_Gowdy_standard} is usually not used.
Instead one takes a harmonic time coordinate $\tau$ that becomes infinite
at the singularity \cite{vanPutten}. That means that even if the difference
of $\tau$ between the space-like slices is fixed,
the difference of $t$ becomes small. Yet, because of our method to
construct the computational mesh we cannot control the slicing in this way.
The results of other numerical methods are thus not really
comparable.

However, simulations of collapsing polarised Gowdy evolutions can be
found e.g. in \cite{Garfinkle:PhysRevD.65.044029,vanPutten}.
Furthermore \cite{Alcubierre:2003pc}
reports about results of the ADM and BSSN codes and the PITT group
\cite{PITTGroup} presents results of the ABIGEL code.
Further information about cosmological space-times with singularities can
be found e.g. in \cite{LivingReviewBerger}.

Numerical simulations that use the expanding polarised Gowdy solution
as a testbed can be found e.g. in \cite{Kim_New:PhysRevD.58.064022}.
Again the PITT group \cite{PITTGroup} presents results of the
ABIGEL code.

Summarising the hitherto existing work about discrete differential
forms in numerical GR it seems that the most urgent task, in order to
apply this method to physically more relevant situations, is to find
statements about the underlying system of equations. Indeed it is not
clear how to apply standard concepts of numerical GR, like
hyperbolicity, to the Cartan formulation.
What is currently lacking is a better
understanding of the notions of a discrete geometry.


\section*{Acknowledgements}
The authors are grateful to A Bossavit for helpful discussions and suggestions.
This work is supported by the SFB 382 project on ''Methods and algorithms for
the simulation of physical processes on high performance computers''.

\appendix
\section{Derivation of the reduced system}
\label{app:wave_symm}

\newcommand{\e}{\mathrm{e}}
\newcommand{\Zb}{\mathbf{Z}}

We are interested in space-times $(\cM,g)$ which can be characterised by the
following conditions:
\begin{enumerate}
\item There is an effective, isometric action of $\RR^2$ on $\cM$.
\item The generators of this action, the two Killing vectors $\xi$ and
  $\eta$, are
  space-like, they are orthogonal and they commute.
\item The action is hypersurface-orthogonal, i.e., the 2-flats
  orthogonal to the span of the Killing vectors are integrable.
\end{enumerate}
From condition~(1) it follows that the space-time is locally foliated
by 2-dimensional orbits whose tangent spaces are spanned by the
Killing vectors. Since these are orthogonal and space-like we can
define two space-like unit vectors by
\begin{align}
\label{eq:def_e2_e3}
\xi &= \e^f e_2, & \eta &= \e^g e_3,
\end{align}
with functions $f$ and $g$ which are constant on the orbits. We
complete these vectors to an orthonormal basis
$(e_0,e_1,e_2,e_3)$ with dual basis
$(\thetab^0,\thetab^1,\thetab^2,\thetab^3)$. Since $[e_2,e_3]=0$ we have
\begin{align}
\dd\thetab^0(e_2,e_3) = e_2(\thetab^0(e_3)) - e_3(\thetab^0(e_2)) -
\thetab^0([e_2,e_3]) = 0
\end{align}
and, similarly, for $\thetab^1$. Hence, we get the expansions
\begin{equation}
  \dd\thetab^0 = \mathbf{A} \thetab^0 + \mathbf{B} \thetab^1,\qquad 
  \dd\thetab^1 = \mathbf{C} \thetab^0 + \mathbf{D} \thetab^1
\label{eq:theta01exp}
\end{equation}
for some 1-forms $\mathbf{A}$, $\mathbf{B}$, $\mathbf{C}$ and
$\mathbf{D}$. This can be expressed by the well known integrability
conditions
\begin{align}
\dd\thetab^0\thetab^0\thetab^1 = 0,\qquad 
\dd\thetab^1\thetab^0\thetab^1 = 0.
\end{align}
While the choice of $e_2$ and $e_3$ is fixed by aligning them with the
Killing vectors the other two basis vectors are not fixed. They can
still rotate inside the plane which they generate with a rotation
which may not only depend on the orbit but even the location on the
orbit. We can fix the latter dependence by requiring that $e_0$ and
$e_1$ commute with both Killing vectors. This requirement is
consistent because the Killing vectors commute.

Then the four basis vectors
are Lie transported along both Killing vectors and the same is true
for the dual basis vectors, i.e., the equations
\begin{align}
\label{eq:frame_invariance}
\cL_\xi \thetab^a = 0 = \cL_\eta \thetab^a
\end{align}
hold for $a=0,\ldots,3$. Using these equations and the
expansions~\eqref{eq:theta01exp} leads us after a short calculation to
the fact that the 1-forms $\mathbf{A}$ to $\mathbf{D}$ are linear
combinations of $\thetab^0$ and $\thetab^1$ only, i.e., they vanish
when restricted to an orbit.

Furthermore, condition~(3) implies that the distribution
generated by $e_0$ and $e_1$ is integrable, i.e., that
\begin{align}
[e_0,e_1] = \alpha e_0 + \beta e_1.
\end{align}
This implies that there exist 2-dimensional submanifolds orthogonal to
the orbits of the action which are dragged into each other by the
group action.

Expressed in terms of the dual basis, the integrability condition is 
\begin{align}
\dd\thetab^2\thetab^2\thetab^3 = 0,\qquad 
\dd\thetab^3\thetab^2\thetab^3 = 0,
\end{align}
which leads to a similar expansion of the differentials in terms of
the coframe as above.
In an analogous calculation making use of the invariance of the frame
\eqref{eq:frame_invariance} 
we find the following relations
\begin{equation}
  \label{eq:relations}
  \dd \thetab^2 = \dd f\, \thetab^2, \quad \dd \thetab^3 = \dd g\,
  \thetab^3. 
\end{equation}
These relationships allow us to partly determine the connection forms from
the first structure equation
\begin{align}
\dd \thetab^a + \omegab^a{}_b \thetab^b = 0.
\end{align}
We find that all the connection forms except for $\omegab:=\omegab^0{}_1$ are
determined from the above equations
\begin{equation}
  \label{eq:connectionforms}
  \omegab^2{}_0 = f_0 \thetab^2, \quad
  \omegab^2{}_1 = f_1 \thetab^2, \quad
  \omegab^3{}_0 = g_0 \thetab^3, \quad
  \omegab^3{}_1 = g_1 \thetab^3, \quad
  \omegab^2{}_3 = 0,
\end{equation}
where $f_0 = e_0(f)$, etc. so that $\dd f = f_0 \thetab^0 + f_1
\thetab^1$. So we find that the information about the extrinsic
geometry of the orbits is contained in the two scalars $f$ and $g$,
while the geometry of the orbit space $\cM_1$ which can be identified
with the space orthogonal to the orbits is described by the coframe
vectors $(\thetab^0,\thetab^1)$ and the connection form $\omegab$.

With these expressions for the connection forms we get the
Nester-Witten 2-forms
\begin{equation}
  \begin{aligned}
    L_0 &= (f_1 + g_1) \thetab^2\thetab^3,&
    L_2 &= -(\star\dd g) \thetab^3 - \omegab \thetab^3,\\
    L_1 &= (f_0 + g_0) \thetab^2\thetab^3,& L_3 &= (\star\dd f) \thetab^2
    + \omegab \thetab^2
  \end{aligned}
\end{equation}
and the Sparling 3-forms
\begin{equation}
\begin{aligned}
  S_0 &= \left( (f_0+g_0) \omegab 
  + f_1 \dd g + f_0 \,(\star\dd g) \right) \thetab^2\thetab^3,&
  S_2 &= \omegab\, \dd g\, \thetab^3,\\
  S_1 &= \left( (f_1+g_1) \omegab
  + f_0 \dd g + f_1\,(\star\dd g) \right) \thetab^2\thetab^3,&
  S_3 &= -\omegab\, \dd f\, \thetab^2.
\end{aligned}
\end{equation}
Here, we have used the notation $\star\alphab$ for the 1-form
$\alpha_0\bo\theta^1 + \alpha_1\bo\theta^0$ given the 1-form
$\alphab=\alpha_0\thetab^0 + \alpha_1 \thetab^1$.  Inserting these
expressions into the equations $\dd L_i = S_i$ (and stripping off the
common factor $\thetab^2\thetab^3$, $\thetab^3$ and $\thetab^2$,
respectively) yields the four equations on $\cM_1$
\begin{align}
\label{eq:Einstein_eqn_1+1}
\begin{gathered}
  \dd (f_1 + g_1) + (f_1 + g_1) (\dd f + \dd g)
  = (f_0+g_0) \omegab + f_1 \dd g + f_0\,(\star\dd g),\\
  \dd (f_0 + g_0) + (f_0 + g_0) (\dd f + \dd g)
  = (f_1+g_1) \omegab + f_0\dd g + f_1\,(\star\dd g),\\
  \dd (\star\dd g) + \dd g(\star\dd g) + \dd \omegab = 0,\\
  \dd (\star\dd f) + \dd f(\star\dd f) + \dd \omegab = 0.
\end{gathered}
\end{align}
The first pair of equations are 1-form equations. They comprise four
scalar equations. However, they contain $\omegab$ which indicates that
they depend on the chosen gauge. We can extract two gauge invariant
equations by multiplying with the basis 1-forms and combining the
equations linearly. This leads to the 2-form equations
\[
\begin{aligned}
  \dd^2 (f + g) + \dd(f+g) \dd(f+g) &= 0,\\
  \dd (\star\dd f)  + \dd (\star\dd g) + \dd(f+g) ((\star\dd f) + (\star\dd g)) &= 0.
\end{aligned}
\]
The first of these is an identity while the second equation can be
combined with the third and fourth equation of
\eqref{eq:Einstein_eqn_1+1}. So we have the following three 2-form
equations
\[
\begin{gathered}
  \dd \omegab = \dd f (\star\dd g),\\
  \dd (\star\dd g) + \dd g(\star\dd g) + \dd f (\star\dd g) = 0,\\
  \dd (\star\dd f) + \dd f(\star\dd f) + \dd g (\star\dd f) = 0.
\end{gathered}
\]

Note, that we do not attach a meaning to the forms $\star \dd f$ and
$\star \dd g$ as yet. These symbols are so far only abbreviations for
the forms $\star \dd f = f_0 \thetab^1 + f_1 \thetab^0$ and $\star \dd
g = g_0 \thetab^1 + g_1 \thetab^0$.

Thus, we end up with the following set of differential forms which we
regard as being defined on the frame bundle over $\cM_1$
\[
\begin{aligned}
  \Zb_0&\equiv
  \dd (f_0 + g_0) + (f_0 + g_0) (\dd f + \dd g)\\
  &- (f_1+g_1) \omegab - (f_0 g_0 + f_1 g_1) \thetab^0 -
  (f_1 g_0 + f_0 g_1) \thetab^1,\\
  \Zb_1&\equiv
  \dd (f_1 + g_1) + (f_1 + g_1) (\dd f + \dd g)\\
  &- (f_0+g_0) \omegab - (f_0 g_0 + f_1 g_1) \thetab^1 -
  (f_1 g_0 + f_0 g_1) \thetab^0,\\
  \Zb_2&\equiv  \dd \omegab - \dd f (\star\dd g),\\
  \Zb_3&\equiv  \dd (\star\dd g) + \dd g(\star\dd g) + \dd f (\star\dd g),\\
  \Zb_4&\equiv  \dd (\star\dd f) + \dd f(\star\dd f) + \dd g (\star\dd f) ,\\
  \Zb_5&\equiv \dd \thetab^0 + \omegab\thetab^1,\\
  \Zb_6&\equiv \dd \thetab^1 + \omegab\thetab^0,\\
  \Zb_7&\equiv \dd f - f_0 \thetab^0 - f_1 \thetab^1,\\
  \Zb_8&\equiv \dd g - g_0 \thetab^0 - g_1 \thetab^1,\\
  \Zb_{9}&\equiv (\dd f_0 - \omegab f_1) \thetab^0 + (\dd f_1 -
  \omegab
  f_0) \thetab^1,\\
  \Zb_{10}&\equiv (\dd g_0 - \omegab g_1) \thetab^0 + (\dd g_1 -
  \omegab
  g_0) \thetab^1,\\
  \Zb_{11}&\equiv \star\dd f - f_0 \thetab^1 - f_1 \thetab^0,\\
  \Zb_{12}&\equiv \star\dd g - g_0 \thetab^1 - g_1 \thetab^0,\\
  \Zb_{13}&\equiv \dd(\star \dd f) - (\dd f_0 - \omegab f_1) \thetab^1
  - (\dd f_1 - \omegab f_0) \thetab^0,\\
  \Zb_{14}&\equiv \dd (\star \dd g ) - (\dd g_0 - \omegab g_1)
  \thetab^1 - (\dd g_1 - \omegab g_0) \thetab^0.
\end{aligned}
\]
Apart from the forms arising from the Witten-Sparling equations this
system includes forms which implement the expansion of $\dd f$, $\dd
g$, $\star\dd f$ and $\star\dd g$ in terms of their components $f_0$,
$f_1$, $g_0$ and $g_1$ as well as their exterior derivatives.

Note, that there is also the first Bianchi identity which should be
added. It reads
\begin{equation}
\dd \omegab\thetab^0 = 0 = \dd \omegab \thetab^1.  
\end{equation}
However, since it is an identity and, in addition, a 3-form equation
which cannot be implemented in our 2-dimensional problem we will
simply assume its general validity without listing it among the
relevant forms.

A closer and somewhat tedious examination of the system reveals the
following facts:
\begin{equation}
\begin{gathered}
  \dd \Zb_0 \in
  \left< \Zb_0, \Zb_1, \Zb_2, \Zb_3, \Zb_4, \Zb_7, \Zb_8, \Zb_{9},
    \Zb_{11}, \Zb_{12}, \Zb_{13} \right>\\
  \dd \Zb_1 \in
  \left<\Zb_0, \Zb_1, \Zb_2, \Zb_3, \Zb_4, \Zb_7, \Zb_8, \Zb_{10},
    \Zb_{11}, \Zb_{12}, \Zb_{14}\right>\\
  \dd \Zb_2 \in \left<\Zb_3\right> \cap \left< \Zb_4 \right>, \quad
  \dd \Zb_3 \in \left< \Zb_3 \right>,\quad  \dd \Zb_4 \in \left<
    \Zb_4 \right>,\\
  \dd \Zb_5  \in \left<\Zb_6 \right>, \quad 
   \dd \Zb_6 \in \left< \Zb_5 \right>,\\
   \dd \Zb_7  \in \left<\Zb_{9} \right>, \quad 
   \dd \Zb_8 \in \left< \Zb_{10} \right>,\quad
   \dd \Zb_{9}  \in \left<\Zb_5,\Zb_6 \right>, \quad 
   \dd \Zb_{10} \in \left<\Zb_5,\Zb_6 \right>,\\%
   \dd \Zb_{11}  \in \left<\Zb_{13} \right>, \quad 
   \dd \Zb_{12} \in \left< \Zb_{14} \right>,\quad
   \dd \Zb_{13}  \in \left<\Zb_5,\Zb_6 \right>, \quad 
   \dd \Zb_{14} \in \left<\Zb_5,\Zb_6 \right>.
\end{gathered}
\end{equation}
The crucial forms are $\Zb_2,\ldots,\Zb_6$. $\Zb_2,\Zb_3,\Zb_4$ determine the
connection and $\Zb_5$ and $\Zb_6$ determine the metric in terms of the
frame $\thetab^0$ and $\thetab^1$. The forms $\Zb_7,\ldots,\Zb_{14}$
are `book keeping' forms in the sense that they express $\dd f$, $\dd
g$, $\star \dd f$ and $\star\dd g$ in terms of the components
$f_0$,$f_1$, $g_0$ and $g_1$. These are not really necessary if we are
only interested in the geometry of $\cM_1$. In fact, in this case it
is not even necessary to know about the functions $f$ and $g$. 

We can extract several systems from these equations. The first system
is obtained by ignoring that $f_0,\dots,g_1$ are components of 1-forms
with respect to the coframe and working simply with these four
functions. The system we use consists of $\Zb_0,\Zb_1,\Zb_5,\Zb_6$, as
well as $\Zb_3+\Zb_2$ and $\Zb_4+\Zb_2$. In addition, we include
$\Zb_9$, i.e., the exterior derivative of $\Zb_7$. The corresponding
equation for $g_0$ and $g_1$ follows from $\Zb_0$ and $\Zb_1$. This is
the system~\eqref{eq:pre_min_system}.

The second system is obtained by eliminating the components as
completely as possible. We use generic 1-forms $\alphab$, $\betab$,
$\gammab$ and $\boldsymbol\delta$ for the 1-forms $\dd f$, $\star \dd
f$, $\dd g$ and $\star\dd g$, respectively. Then the system consists
of $\Zb_0\thetab^0 + \Zb_1 \thetab^1$, linear combinations of $\Zb_2$,
$\Zb_3$ and $\Zb_4$ as well as $\Zb_5$ and $\Zb_6$. Finally we add the
closedness of $\alphab$ and $\gammab$ and the algebraic relationships
between $\alphab$ and $\betab$ respectively $\gammab$ and
$\bo\delta$. This results in the system~\eqref{eq:grav_wave_first_deg}.

Note, that it is enough to include the combination of $\Zb_0$ and $\Zb_1$
given above. This is due to the following consideration. Let
$I=\left<\Zb_k, 2\le k \le 14\right>$ be the ideal generated by all
forms except for $\Zb_0$ and $\Zb_1$. Then we have
\begin{equation}
\begin{aligned}
  \dd \Zb_0 - \omegab\, \Zb_1 + \dd(f+g) \Zb_0 &\equiv 0 \mod I,\\
  \dd \Zb_1 - \omegab\, \Zb_0 + \dd(f+g) \Zb_1 &\equiv 0 \mod I.
\end{aligned}
\end{equation}
By taking another exterior derivative we get the equations
\begin{equation}
\begin{aligned}
  \dd \omegab\, \Zb_1 &\equiv 0 \mod I,\\
  \dd \omegab\, \Zb_0 &\equiv 0 \mod I,
\end{aligned}
\end{equation}
since $\dd I \subset I$. This implies, that if all other equations
hold, then we have $\dd\omegab\, \Zb_i=0$ which forces $\Zb_i=0$ unless
$\dd\omegab=0$, a case in which we are not interested.



\end{document}